\documentclass[12pt]{iopart}
\usepackage{graphics,epsfig}

\def\be{\begin{equation}}
\def\ee{\end{equation}}
\def\bea{\begin{eqnarray}}
\def\eea{\end{eqnarray}}

\def\frac#1#2{{\textstyle{#1\over#2}}}
\def\darr#1{\raise1.5ex\hbox{$\leftrightarrow$}\mkern-16.5mu #1}
\def\){\right)}
\def\({\left( }
\def\]{\right] }
\def\[{\left[ }
\def\si{{}^1\kern-.14em S_0}
\def\siii{{}^3\kern-.14em S_1}
\def\diii{{}^3\kern-.14em D_1}

\def\cpt{\chi PT}

\def\vkay{{\vec k}}
\def\vkayprime{{{\vec k}\, '}}
\def\vpee{{\vec p}}
\def\vpeeprime{{{\vec p}\, '}}

\def\lsim{\mathrel{\rlap{\lower4pt\hbox{\hskip1pt$\sim$}}        
    \raise1pt\hbox{$<$}}}         
\def\gsim{\mathrel {\rlap{\lower4pt\hbox{\hskip1pt$\sim$}}
    \raise1pt\hbox{$>$}}}         

\begin{document}

\title[$\chi$PT for electroweak reactions on deuterium]
      {Chiral perturbation theory for 
electroweak reactions on deuterium}

\author{Daniel R. Phillips}
\address{Department of Physics and Astronomy, Ohio University, 
	Athens, OH 45701, USA}

\begin{abstract}
I summarize two recent applications of chiral perturbation theory to
electromagnetic reactions on deuterium: elastic electron-deuteron
scattering, and Compton scattering on deuterium. Both calculations
have now been carried out to three orders in the chiral expansion. The
expansion shows good convergence and is able to reproduce data for
$|{\bf q}| \lsim 600$ MeV in $ed$ and for $\omega=55$-- 95 MeV in
$\gamma d$. These results demonstrate that $\chi$PT can be used to
reliably compute operators and wave functions for
low-momentum-transfer reactions in light nuclear systems.
\end{abstract}

\section{Introduction}

In recent years effective field theory (EFT) techniques have found
increasing acceptance in nuclear physics.  EFT facilitates the
systematic separation of the effects of high-momentum physics from
those of low-momentum physics. In nuclear physics there are many
different momentum scales present. But for the $A=2$ system the
key ones would seem to be:
\begin{equation}
\sqrt{MB} \ll m_\pi \ll \Lambda_{\chi {\rm SB}},
\end{equation}
with $B$ the binding energy of deuterium, $m_\pi$ and $M$ the pion and
nucleon masses, and $\Lambda_{\chi \rm{SB}}$ the scale of
chiral-symmetry breaking, which is of the order of the mass of the $\rho$
meson. An EFT based on the first hierarchy---$\sqrt{MB} \ll
m_\pi$---has had considerable success describing very-low-energy
reactions in $A=2$ and $A=3$ (for reviews see~\cite{vK99,Be00,BvK01}),
and has recently been extended to $A=4$~\cite{Ha04}. Here I will focus
on the EFT built on the second hierarchy---$m_\pi \ll \Lambda_{\chi
\rm{SB}}$. This EFT is chiral perturbation theory ($\chi$PT)~(for a
review see \cite{Br95}). Here the low-momentum physics is described by
a quantum field theory of nucleons and pions whose Lagrangian is the
most general one that respects the approximate chiral symmetry of QCD
and the pattern of its breaking.  Higher-energy effects of QCD appear
in $\chi$PT as non-renormalizable contact operators.  Any process
involving nucleons and an arbitrary number of (soft) pions and photons
should then be computable as an expansion in the ratio of nucleon or
probe momenta (denoted here by $p$ and $q$) and $m_\pi$ to
$\Lambda_{\chi \rm{SB}}$.

Since typical momenta in light nuclei are $\lsim m_\pi$ we would
expect to be able to calculate the response of such nuclei to
low-momentum electroweak probes using $\chi$PT. As we will see below,
$\chi$PT provides a systematic way to compute what in
nuclear-structure physics would be called ``bare operators''.  But the
operators computed in this fashion include pieces which account for
the fact that they are not actually ``bare'' if one thinks about the
physics at scale $\Lambda_{\chi {\rm SB}}$ which has been ``integrated
out'' of the low-momentum theory of nucleons and pions: high-momentum
nucleon modes, the delta isobar, mesons such as the rho, pomerons,
M-branes, etc.

Here I will argue that these operators produce results for
electromagnetic few-nucleon reactions which are model independent, can
be systematically improved by going to higher order, and yield a good
description of the data over the region of validity of the EFT.
Section 2 outlines the $\chi$PT expansion, and briefly discusses how
to obtain deuteron wave functions using it.  Section 3
then looks at electron-deuteron scattering in $\chi$PT as a probe of
deuteron structure. Section 4 examines the use of Compton scattering on
the deuteron---in particular as a way to extract isoscalar
polarizabilities of the nucleon.

\section{Power counting and deuteron wave functions}

\subsection{Power counting}

Consider an elastic scattering process on the deuteron.  If $\hat{O}$
is the transition operator for this reaction then the amplitude in
question is simply 
$\langle \psi| \hat{O} |\psi \rangle$,
with $|\psi \rangle$ the deuteron wave function. In this section, we
follow Weinberg~\cite{weinberg}, and divide the formulation of a
systematic expansion for this amplitude into two parts: the expansion
for $\hat{O}$, and the construction of $|\psi \rangle$.

Chiral perturbation theory gives a systematic expansion
for $\hat{O}$ of the form
\begin{equation} 
\hat{O}=\sum_{n=0}^\infty \hat{O}^{(n)},
\label{eq:expansion}
\end{equation} 
where we have labeled the contributions to $\hat{O}$ by their order
$n$ in the small parameter $P \equiv p/\Lambda_{\chi {\rm SB}}$,
$q/\Lambda_{\chi {\rm SB}}$, $m_\pi/\Lambda_{\chi {\rm SB}}$.

To construct $\hat{O}^{(n)}$ one first writes down
the vertices appearing in the chiral Lagrangian up to order $n$. One
then draws all of the two-body, two-nucleon-irreducible, Feynman
graphs for the process of interest which are of chiral order $P^n$. The
rules for calculating the chiral order of a particular graph are:

\begin{itemize}
\item Each nucleon propagator scales like $1/P$ (provided that
the energy flowing through the nucleon line is $\sim m_\pi$);

\item Each loop contributes $P^4$;

\item Graphs in which both particles participate in the reaction
acquire a factor of $P^3$;

\item Each pion propagator scales like $1/P^2$;

\item Each vertex from the $n$th-order piece of the chiral Lagrangian
contributes $P^n$.  
\end{itemize}
In consequence more complicated contributions involving
two-body mechanisms, and/or higher-order vertices, and/or more loops
are suppressed by powers of $P$.

\subsection{Deuteron wave functions}

This leaves us with the problem of constructing a deuteron wave
function that is consistent with the operator $\hat{O}$. Weinberg's
proposal was to use the $\chi$PT expansion (\ref{eq:expansion}) for
the $NN$ potential $V$, and then solve the Schr\"odinger equation to
find the deuteron (or other nuclear) wave
function~\cite{weinberg}. Matrix elements $\langle \psi|\hat{O}|
\psi \rangle$ should then incorporate the consequences of
chiral-symmetry breaking in a model-independent way.

Calculations have shown that the $NN$ phase shifts can be understood,
and deuteron bound-state static properties reliably computed, with
$NN$ potentials derived from
$\chi$PT~\cite{Or96,KW97,Ep99,Re99,EM03}. This is now a sophisticated
enterprise, with N$^3$LO $[$O$(P^4)]$ potentials recently having been
obtained~\cite{EM03,Ep04,Ma04}. These potentials can reproduce $NN$
data with an accuracy which rivals that of ``high-quality'' $NN$
potentials, at least for $T_{\rm lab} < 200$ MeV.

In the Goldstone-boson and single-nucleon sector $\chi$PT works
because loop effects are generically suppressed by powers of the small
parameter $P$. In zero and one-nucleon reactions the power counting in
$P$ applies to the amplitude, and not to the two-particle
potential. However, the existence of nuclei tells us immediately that
a power counting in which all loop effects are suppressed cannot be
correct for the multi-nucleon case, since if it were there would be no
nuclei, and neither we, or this workshop, would occur. Weinberg's
proposal to instead power-count the potential is one response to this
dilemma. However, its consistency has been vigorously debated in the
literature (see~\cite{vK99,Be00,BvK01} for reviews).  Beane {\it et
al.}~\cite{Be01} have shown that Weinberg's proposal is consistent to
leading order in the $\siii-\diii$ channel, but there remain questions
about the consistency in other channels and at higher
orders~\cite{Ka96,No04}.  Here we will use wave functions calculated
according to the Weinberg proposal. One of our objectives will be to
assess whether such a procedure really does lead to convergent EFT
expansions for the processes under consideration, or whether problems
along the lines of those raised in Refs.~\cite{Ka96,No04} also occur
in the chiral EFT for electroweak reactions on deuterium.

\section{Elastic electron scattering on deuterium}

One quantitative test of $\chi$PT's deuteron wave functions is
provided by elastic electron-deuteron scattering. We thus turn our
attention to the deuteron electromagnetic form factors $G_C$, $G_Q$,
and $G_M$. These are matrix elements of the deuteron current $J_\mu$,
with:
\begin{eqnarray}
\qquad G_C=\frac{1}{3 |e|} \left(\left \langle 1\left|J^0\right|1 \right \rangle + 
\left \langle 0\left|J^0\right|0 \right \rangle + \left \langle -1\left|J^0\right|-1 \right \rangle
\right),\\ 
G_Q=\frac{1}{2 |e| \eta M_d^2} \left(\left \langle 0\left|J^0\right|0 \right \rangle - \left \langle 1\left|J^0\right|1 \right \rangle\right), \quad
G_M=-\frac{1}{\sqrt{2 \eta}|e|}
\left \langle1\left|J^+\right|0\right \rangle,
\end{eqnarray}
where we have labeled these (non-relativistic) deuteron states by the
projection of the deuteron spin along the direction of the momentum
transfer ${\bf q}$ and $\eta \equiv |{\bf q}|^2/(4 M_d^2)$. $G_C$,
$G_Q$, and $G_M$ are related to the experimentally-measured $A$, $B$,
and $T_{20}$ in the usual way, with $T_{20}$ being primarily sensitive
to $G_Q/G_C$ and $B$ depending only to $G_M$. Here we will compare
calculations of the charge and quadrupole form factor with the recent
extractions of $G_C$ and $G_Q$ from data~\cite{Ab00B}.

\begin{figure}[htbp]
\vspace{0.2cm}
\hspace{-0.25in}
\centerline{\epsfig{figure=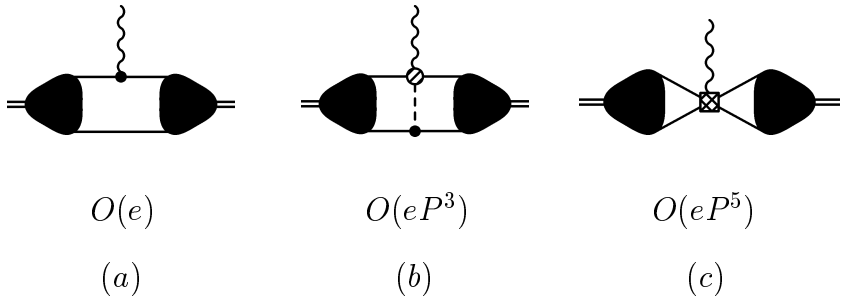,width=10.0cm}}
\vspace{-0.1cm}
\caption{Diagrams representing the leading contribution to 
$J_0$ [(a)], the leading two-body contribution [(b)], and the dominant
short-distance piece [(c)].  Solid circles are vertices from ${\cal
L}_{\pi N}^{(1)}$, and the shaded circle is the vertex from ${\cal
L}_{\pi N}^{(2)}$.}
\label{deuterongraphs}
\end{figure}

Both of these form factors involve the zeroth-component of the
deuteron four-current $J^0$. Here we split $J^0$ into two pieces: a
one-body part, and a two-body part. The one-body part of $J^0$ begins
at order $|e|$ (the proton charge) with the impulse approximation
diagram calculated with the non-relativistic single-nucleon charge
operator for structureless nucleons. Corrections to the
single-nucleon charge operator from relativistic effects and nucleon
sub-structure are suppressed by two powers of $P$, and thus arise at
$O(e P^2)$, which is the next-to-leading order (NLO) for $G_C$ and
$G_Q$. At this order one might also expect meson-exchange current
(MEC) contributions. However, all MECs constructed with vertices from
${\cal L}_{\pi N}^{(1)}$ are isovector. These play a role in, e.g. $np
\rightarrow d \gamma$~\cite{Pa95}.  The first MEC effect in $ed
\rightarrow ed$ does not occur until N$^2$LO, or $O(e P^3)$, where an
$NN \pi \gamma$ vertex from ${\cal L}_{\pi N}^{(2)}$ gets included in
an isoscalar two-body contribution to the deuteron charge operator
(see Fig.~\ref{deuterongraphs}(b))~\footnote{This exchange-charge
contribution was first derived by Riska~\cite{Ri84}.}.

The most important correction that arises at NLO is the inclusion of
{\it nucleon} sub-structure. At $O(e P^2)$ the isoscalar nucleon form
factors are dominated by short-distance physics, and so the only
correction to the point-like leading-order result comes from the
inclusion of the nucleon's electric radius. For the isoscalar
combination of nucleon electric form factors $\chi$PT to $O(e P^2)$
gives:
\begin{equation}
{G_E^{(s)}}_{\mbox{$\chi$PT NLO}}=1 - \frac{1}{6} \langle r_E^{(s) \, 2}
\rangle q^2.
\end{equation}
This description of nucleon structure breaks down at momentum
transfers $q$ of order 300 MeV. There is a concomitant failure in the
description of $ed$ scattering data~\cite{MW01,Ph03}.  In order to
focus on {\it deuteron} structure, in the results presented below I
have chosen to circumvent this issue by using a ``factorized''
inclusion of nucleon structure~\cite{Ph03}: $\chi$PT is used to
compute the ratio $\frac{G_C}{G_E^{(s)}}$. This allows us to use
experimentally-measured single-nucleon form factors~\footnote{There is
a bit of an issue of circularity here, since $ed$ scattering data is one
input to the extraction of the neutron electric form factor.} in the
calculation, thereby allowing us to test how far the theory is able to
describe the $NN$ dynamics.

\begin{figure}[thbp]
\begin{center}
\epsfig{figure=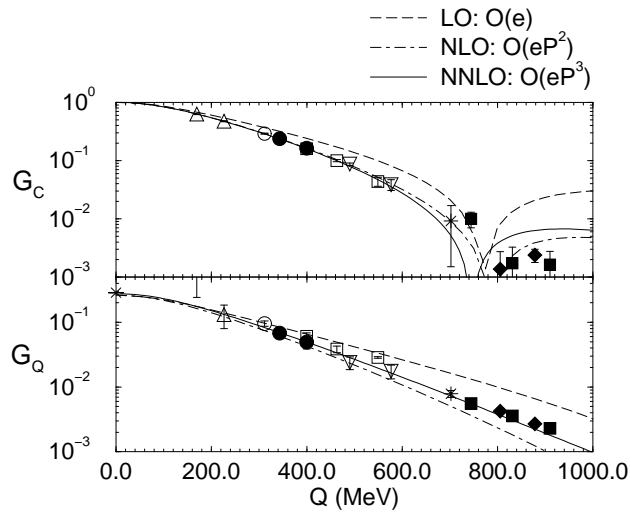,height=0.3\textheight}
\end{center}
\caption{The
deuteron charge and quadrupole form factors to NNLO in chiral
perturbation theory. The experimental data is taken from the
compilation of Ref.~\cite{Ab00B}. $G_Q$ is in units of fm$^2$.}
\label{fig-FCFQ}
\end{figure}

The results for $G_C$ and $G_Q$ are shown in Fig.~\ref{fig-FCFQ}. The
figure demonstrates that convergence is quite good below $q \sim 700$
MeV. The results shown are for the NLO chiral wave function.  It is
clear that---provided information from single-nucleon form factors is
taken into account---$\chi$PT is perfectly capable of describing the
charge and quadrupole form factors of deuterium at least as far as the
minimum in $G_C$. 

$G_M$ can be obtained in a similar way.  The LO
contribution to $G_M$ is $O(e P)$. The first two-body mechanisms
enter at $O(e P^4)$, when an undetermined two-body counterterm, as well
as a long-range two-body current with an undetermined coefficient,
appear~\cite{MW01,Pa99}. Results for $G_M$ at $O(e P^3)$ turn out to be
of similar quality to those for $G_C$~\cite{Ph03}, but are
somewhat more sensitive to short-distance physics, as expected
given the presence of a counterterm at NNLO in this observable.

Using the NNLO chiral wave function, high-quality potential-model wave
functions, or indeed simple wave functions which include only one-pion
exchange regulated by a square well at short distances, does not
modify the plots for $G_C$ and $G_Q$ greatly below $q=700$
MeV~\cite{Ph03}. As long as wave functions with the same long-distance
parts are employed these observables are apparently not very sensitive
to the choice of $|\psi \rangle$. In the language of the no-core shell
model they do not ``renormalize strongly''.  (An important exception
is the deuteron quadrupole moment, $Q_d$.)  This is good news for
Weinberg's power counting, because the first operators with which
$J_0$ can be renormalized occur at $O(e P^5)$
(Fig.~\ref{deuterongraphs}(c)). Therefore it was to be expected that
operator renormalization would have this small impact on $G_C$ and
$G_Q$. Furthermore, the dominant piece of the operator
renormalization can be taken into account by adjusting the
coefficients of the $O(e P^5)$ operators to reproduce, e.g. $\langle
r_d^2 \rangle$ and $Q_d$. This would then push the omitted effects of
operator renormalization in these observables to $O(e P^7)$.

\section{Compton scattering on deuterium}
\label{sec-gammagammad}

\noindent
Compton scattering on the nucleon at low energies is a fundamental
probe of the long-distance structure of these hadrons.  This process
has been studied in $\cpt$ in \cite{Br92,Br92B}. The
following results for the proton polarizabilities were obtained at $O(e^2 P)$:
\begin{equation}
\alpha_p={{5 e^2 g_A^2}\over{384 \pi^2 f_\pi^2 m_\pi}} =12.2 \times
10^{-4} \, {\rm fm}^3, \quad \beta_p={{1} \over {10}} \alpha_p= 1.2 \times
10^{-4} \, {\rm fm}^3.
\end{equation}
Going to next order in the chiral expansion 
single-nucleon counterterms which shift the
polarizabilities enter the calculation. Using the $\gamma p$ amplitude
computed by McGovern to $O(e^2 P^2)$~\cite{judith} we fitted $\gamma p$ data
over the kinematic range $\omega, \sqrt{|t|} < 180$ MeV~\cite{Be02}.
\begin{figure}[htb]
   \epsfxsize=10.0cm
   \centerline{\epsffile{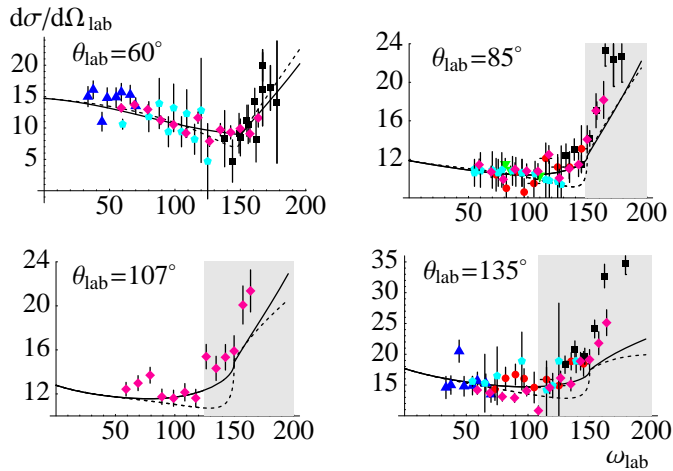}}
\caption{\label{Q4deuprotfit} Results of the $O(e^2 P^2)$ EFT best fit to
   the differential cross sections for Compton scattering on the
   proton at various angles, compared to the experimental data
   from various facilities. The gray region is excluded from the
   fit. References for the data and a legend for the symbols are
   given in Ref.~\cite{judith}.}
\end{figure}

The best fit for the proton electric and
magnetic polarizabilities is 
\begin{equation}
\alpha_p = (12.1 \pm 1.1)_{-0.5}^{+0.5} \times 10^{-4} \, {\rm fm}^3, \quad
\beta_p = (3.4 \pm 1.1)_{-0.1}^{+0.1} \times 10^{-4} \, {\rm fm}^3. 
\label{eq:protpol1}
\end{equation}
Statistical (1-$\sigma$) errors are inside the brackets. An estimate
of the theoretical error due to truncation of the expansion at $O(e^2
P^2)$ is given outside the brackets. The result (\ref{eq:protpol1}) is
compatible with other extractions, although the central value of
$\beta_p$ is higher~\cite{PDG}.

One would like to perform a similar analysis for $\gamma d$ scattering
data.  Coherent Compton scattering on a deuteron target has been
measured at $E_\gamma=$ 49 and 69 MeV by the Illinois group
\cite{Lu94}, at $E_\gamma= 84.2-104.5$ MeV at Saskatoon~\cite{Ho00},
and at $E_\gamma=55$ and $66$ MeV at MAXlab at Lund~\cite{Lu02}.  Such
data should permit an extraction of the isoscalar nucleon
polarizabilities $\alpha_N$ and $\beta_N$. These are interesting, not
only as fundamental nucleon-structure parameters, but also because
$\chi$PT predicts $\alpha_N=\alpha_p$, $\beta_N=\beta_p$ at $O(e^2
P)$. The amplitude for $\gamma d$ scattering
involves mechanisms other than Compton scattering on the individual
constituent nucleons. So, our desire to extract nucleon
polarizabilities argues for a theoretical calculation of this reaction
that is under control in the sense that it
accounts for {\it all} mechanisms to a given order in $\cpt$.

\begin{figure}[tbhp]
\vskip -2.2cm
\centerline{\epsfig{figure=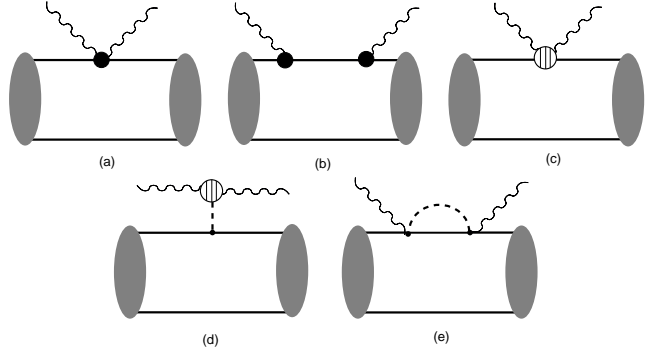,height=.3\textheight}}
\caption{Graphs
   which contribute to Compton scattering on the deuteron at
   ${\cal O}(e^2)$ (a) and ${\cal O}(e^2 P)$ (b-e).  The
   sliced blob in graph (c) is from ${\cal L}_{\pi N}^{(3)}$.
   Crossed graphs are not shown.}
\label{procomptonebodmod} 
\end{figure}

\begin{figure}[t,h.b,p]
\centerline{\epsfig{figure=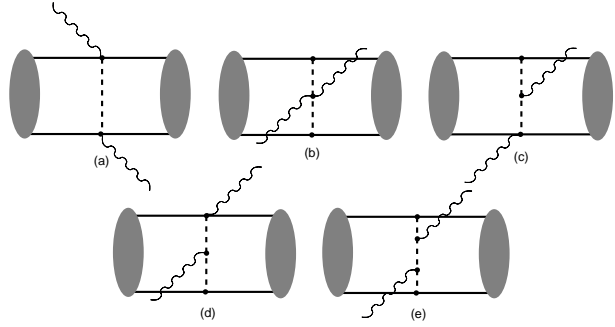,height=.18\textheight}}
\caption{Two-body graphs which contribute to $\gamma d$ scattering at
${\cal O}(e^2 P)$. Crossed graphs are not shown.}
\label{combined} 
\end{figure}

The Compton amplitude we wish to evaluate is (in the $\gamma d$
center-of-mass frame):
\begin{eqnarray}
T^{\gamma d}_{M' \lambda' M \lambda}(\vkayprime,\vkay)&=& \int
\frac{d^3p}{(2 \pi)^3} \, \, \psi_{M'}\left( \vpee + \frac{\vkay -
\vkayprime}{2}\right) \, \, T_{\gamma N_{\lambda'
\lambda}}(\vkayprime,\vkay) \, \, \psi_M(\vpee)\nonumber\\ 
&+& \int \frac{d^3p \, \, d^3p'}{(2 \pi)^6} \, \, \psi_{M'}(\vpeeprime) \, \,
T^{2N}_{\gamma NN_{\lambda' \lambda}}(\vkayprime,\vkay) \, \, \psi_M(\vpee)
\label{eq:gammad}
\end{eqnarray}
where $M$ ($M'$) is the initial (final) deuteron spin state, and
$\lambda$ ($\lambda'$) is the initial (final) photon polarization
state, and $\vkay$ ($\vkayprime$) the initial (final) photon
three-momentum, which are constrained to
$|\vkay|=|\vkayprime|=\omega$.  The first integral is 
represented by the graphs of Fig.~\ref{procomptonebodmod} 
where the photon interacts with only one nucleon.
The second integral corresponds to the graphs of 
Fig.~\ref{combined} where the current is of a two-body nature.

The LO contribution to Compton scattering on the deuteron is shown in
Fig.~\ref{procomptonebodmod}(a).  This graph involves a vertex from
${\cal L}_{\pi N}^{(2)}$ and so is $O({e^{2}})$.  This contribution is
simply the Thomson term for scattering on the proton. There is thus no
sensitivity to either two-body contributions {\it or} nucleon
polarizabilities at this order.  At $O({e^{2}P})$ there are several
more graphs with a spectator nucleon
(Figs.~\ref{procomptonebodmod}(b),(c),(d)), as well as graphs
involving an exchanged pion with leading order vertices
(Fig.~\ref{combined}) and one-loop graphs with a spectator nucleon
(Fig.~\ref{procomptonebodmod}(e))~\cite{Be99}. Graphs such as
Fig.~\ref{procomptonebodmod}(e) contain the physics of the proton and neutron
polarizabilities at $O(e^2 P)$ in $\chi$PT.

\begin{figure}[!ht]
\centerline{\epsfig{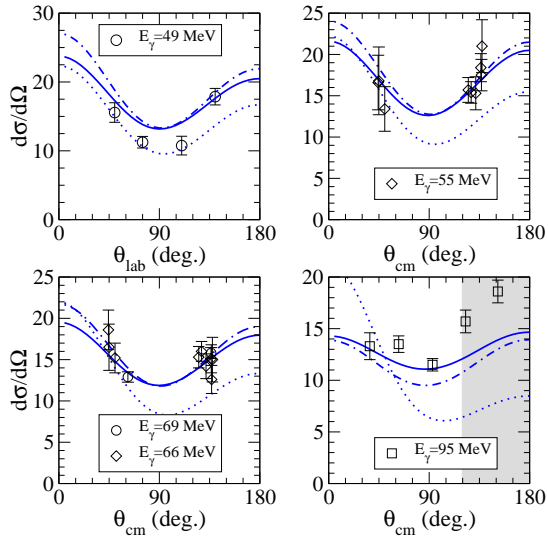}}
\caption{Results of the $O(e^2)$ (dotted), $O(e^2 P)$ (dashed), and
$O(e^2 P^2)$ (solid) calculations for $\gamma d$ scattering as
compared to data at $E_\gamma=49~{\rm MeV}$~\cite{Lu94}, $55~{\rm
MeV}$~\cite{Lu02}, $66$ \& 69 $~{\rm MeV}$~\cite{Lu94,Lu02}, and
$95~{\rm MeV}$~\cite{Ho00}.  Only statistical errors on the data are
shown. The gray region is excluded from the fit.}
\label{cvgplot}
\end{figure}

We employed a variety of wave functions $|\psi \rangle$, and found
moderate wave-function sensitivity. Results shown here are generated
with the NLO chiral wave function of Ref.~\cite{Ep99}.
Fig.~\ref{cvgplot} shows the results of a fit to the world's modern
$\gamma d$ data.  The dotted line is the prediction at $O(e^2)$ in the
kernel, where the second contribution in Eq.~(\ref{eq:gammad}) is
zero, and the single-scattering contribution is given solely by
Fig.~\ref{procomptonebodmod}(a).  It is quite remarkable how well the
$O(e^2)$ calculation reproduces the 49 MeV data. However, the
agreement is somewhat serendipitous: there are significant $O(e^2 P)$
corrections.  At these lower photon energies Weinberg power counting
begins to break down, since it is designed for $\omega \sim m_\pi$,
and does {\it not} recover the deuteron Thomson amplitude as $\omega
\rightarrow 0$.  Correcting the power counting to remedy this
difficulty appears to improve the description of the 49 MeV data,
without significantly modifying the higher-energy
results~\cite{Be02,Be99}.

At $O(e^2 P)$ all contributions to the kernel are fixed in terms of
known pion and nucleon parameters, so to this order $\cpt$ makes {\it
predictions} for deuteron Compton scattering. Note that the $O(e^2 P)$
corrections get larger as $\omega$ is increased---as expected.  The
agreement of the $O(e^2 P)$ calculation with the intermediate-energy
data sets is very good, although only limited conclusions can be
drawn. These results are not very different from other,
potential-model, calculations~\cite{Wl95,LL98,KM99}.

At $O(e^2 P^2)$ terms enter the single-nucleon amplitude which allow
us to fit $\alpha_N$ and $\beta_N$ to $\gamma d$ data. A number of new
two-body currents also appear.  However, there are still no
short-distance $\gamma NN$ operators contributing to $\gamma d$
scattering at this order.  Therefore an $O(e^2 P^2)$ calculation
allows us to test the single-nucleon physics which is used to predict
the results of coherent scattering on deuterium, since there are no
undetermined parameters in the two-body mechanisms that enter.

Unfortunately the data is fairly sparse, and the sensitivity to the
$NN$ wave function quite large, so our fit does not tie down
$\alpha_N$ and $\beta_N$ very accurately. The result is:
\begin{equation}
\alpha_N = (13.0 \pm 1.9)_{-1.5}^{+3.9} \times 10^{-4} \, {\rm fm}^3,
\quad
\beta_N = (-1.8 \pm 1.9)_{-0.9}^{+2.1} \times 10^{-4} \,
{\rm fm}^3.
\label{eq:Npols}
\end{equation}
The errors inside the brackets are statistical and those outside are
an estimate of the effect from higher-order terms. (For details see
Ref.~\cite{Be02}.) The fit only includes data satisfying
$\omega,\sqrt{|t|} < 160$ MeV.  Recent work suggests that to describe
the backward-angle points at 95 MeV requires the inclusion of an
explicit delta degree of freedom~\cite{Hi04}.

The second error quoted on $\alpha_N$ and $\beta_N$ is large because,
as alluded to above, cross sections depend on the choice of $|\psi
\rangle$ at the 10--20\% level. This change in the $\gamma d$ cross
section has a sizable impact on the extraction of $\alpha_N$ and
$\beta_N$.  While not all of the wave functions we employed are
consistent with $\chi$PT, they do all have the correct long-distance
behavior. The differences we see are thus a consequence of different
short-range behavior in, say, the Nijm93 and NLO $\chi$PT $NN$
potentials. As such they should be renormalized by $\gamma NN$ contact
operators. However, Weinberg power counting predicts that the
operators which do this do not appear until (at least) $O(e^2 P^3)$.
Thus the degree of variability seen in our results with different
wave functions could be viewed as inconsistent with
this power counting.  Indeed, it may be necessary to modify the power
counting so that, e.g. $\gamma NN$ contact operators appear at a lower
order than is indicated by the naive $\chi$PT power counting we have
used for $\hat{O}$.  The issue here is one of relevance to
nuclear-structure physics: it speaks of the need to properly
renormalize the effective $\gamma NN$ operators found when high-energy
degrees of freedom are integrated out of our EFT of the $NN$ system.
Further understanding of this issue is also crucial if
$\chi$PT is to be an accurate calculational tool for low-energy
reactions on deuterium.

\section{I thought you said electro{\it weak}?}

There is nothing in what I have presented so far that is particular to
the photon as a probe. Analogous expansions can be made for axial
currents. The result is a systematic approach to reactions such as $pp
\rightarrow d + e^+ + \nu_e$ and $\nu d \rightarrow \nu d$. I do not
have room to properly describe the beautiful work on these reactions
by Park, Ando, and collaborators~\cite{Pa03,An03}. In my view a
particular strength of that work is the application of the same
$\chi$PT electroweak operators to $A=2$, 3, and 4. Ref.~\cite{Pa03}
uses the well-known tritium beta decay rate to fix the strength of an
(otherwise poorly constrained) two-body contribution to the
Gamow-Teller transition operator. This renormalizes the effective $NN$
Gamow-Teller operator to such high accuracy that the $pp$-fusion $S$ factor has
an estimated error from higher-order terms of less than $0.5$\%. The
same one- and two-nucleon operators are then used to predict the
reaction $p + {}^3~{\rm He} \rightarrow {}^4~{\rm He} + e^+ +
\nu_e$~\cite{Pa03} and $\nu d$ scattering~\cite{An03}. This is a great
example of $\chi$PT's ability to give a systematic derivation of
electroweak operators that can be used in a variety of light nuclear
systems to yield accurate predictions.

\section*{Acknowledgments}
I thank the organizers of the `Workshop on Microscopic Nuclear
Structure Theory' for their invitation to participate in a stimulating
and timely meeting.  I am also grateful to Silas Beane, Manuel
Malheiro, Judith McGovern, and Bira van Kolck for a profitable and
enjoyable collaboration.

\end{document}